\documentstyle[prl,aps]{revtex}

\draft
\tighten
\twocolumn

\begin{document}

\title{Modeling the actinides with disordered local moments}

\author{Anders~M.~N.~Niklasson$^{1}$, John M. Wills$^{1}$, Mikhail I. Katsnelson$^{2,3}$, 
Igor A. Abrikosov$^{3}$, Olle Eriksson$^{3}$, and B\"orje Johansson$^{3,4}$}

\address{
$^1$ Theoretical Division, Los Alamos National Laboratory,
Los Alamos, NM 87545, USA\\
$^2$ Institute of Metal Physics, 620219 Ekaterinburg, Russia\\
$^3$ Department of Physics, Uppsala University, Box 530, SE-75121 Uppsala, Sweden \\
$^4$ Applied Materials Physics, Department of Materials Science and Engineering,
Royal Institute of Technology, SE-10044 Stockholm, Sweden}

\date{\today}
\maketitle

\begin{abstract}
{A first-principles disordered local moment (DLM) picture 
within the local-spin-density and coherent potential
approximations (LSDA+CPA) of the actinides is presented. The parameter free 
theory gives an accurate description of bond lengths and bulk modulus.
The case of $\delta$-Pu is studied in particular
and the calculated density of states is compared
to data from photo-electron spectroscopy. The relation between
the DLM description, the dynamical mean field
approach and spin-polarized magnetically ordered
modeling is discussed.
\\
}
\end{abstract}

\section{Introduction}
The elemental actinide metals, Pu in particular, exhibit
several unique features. They are among the most complex
elements in nature, with a rich set of allotropes\cite{Review}, of which
several have complex low symmetry
crystal structures \cite{Soderlind95}. Pu, for example,
is the only element with seven condensed matter phases
at zero pressure of which one (the $\delta$-phase) demonstrates 
negative thermal expansion.
The understanding of these anomalous properties is a serious challenge.
Several of the main features of the actinides can be understood 
in terms of the progressive filling of the 5f shell.
In the light actinides, Th to Np,
the 5f electrons are itinerant and participate in the bonding,
whereas in the heavier actinides, Am to Cf, the 5f electrons
are localized and exhibit behavior more similar to the lanthanides. 
Plutonium takes a particular place between these extremes.
The low temperature $\alpha$-phase has been shown to
be well described with itinerant bonding 5f electrons
whereas the high temperatures phases with their increased
volumes suggests a localized or partly localized configuration.
This behavior indicates a Mott-like transition of the f-electrons 
similar to the $\alpha$-$\gamma$ transition in cerium \cite{johan}.

There have been several first principles approaches to calculating
the properties of the actinides 
\cite{Soderlind95,Skriver78,ReviewB,Solovyev91,Soderlind00,Wills92,Katsnelson92,Ek93,Soderlind94,Olle99,Wang00,Soderlind01}. 
Although the light actinides are well described within density functional theory,  
the local density approximation (LDA) fails in general to describe the 5f localization 
which occurs in actinide compounds, elemental Pu at high temperature, and in the actinides 
past Pu. A number of techniques have therefore been developed to include correlation effects 
beyond LDA in order to describe the partially localized 5f electronic structure.
These include LDA+U\cite{Anisimov91,Bouchet00}, orbital polarization\cite{Soderlind01}, 
and the Mixed Level Model \cite{Olle99}. Of particular relevance to the work presented 
here are recent attempts to combine LDA and Dynamical Mean-Field Theory (DMFT).
DMFT has been shown to be a very useful approximation at the consideration
of strongly correlated electron systems (for review, see \cite{GKKR}).
LDA combined with the DMFT may provide a first principles technique 
to study correlated electron materials, and there have been 
several attempts to apply implementations of LDA+DMFT \cite{anis,LDA++}; 
for a review, see \cite{LK,Held}. Recently this LDA+DMFT approach has been
applied to the Pu problem \cite{savr}. The LDA+DMFT approach gives an opportunity
to describe correlation effects on the electronic structure and properties of $d$- 
and $f$- electron systems. However, the technique is cumbersome and it is not
{\it completely ab initio} because of the problem with the choice of $U$
(see, e.g. \cite{LK}). In addition, implementations of DMFT are computationally
intensive which makes the calculation of complex structures difficult.
Sometimes, because of the complexity of the calculations, uncontrollable 
approximations are made, e.g., using a single $U$ parameter instead of 
a complete interaction matrix \cite{Ce1,Ce2,savr}.

It is commonly accepted that accounting for Hubbard correlation effects is of
crucial importance for $f$-electron systems. 
On the other hand, some of these effects can be taken into account, in an approximate way, 
in the framework of more traditional density functional techniques. 
This is the approach taken in the present study 
where we have modeled the actinides by means of Disordered Local Moments (DLM) \cite{gyorffy}
within the Local Spin Density Approximation and the Coherent Potential Approximation \cite{CPA} (LSDA+CPA).
The purpose of the present work is to investigate and possibly demonstrate that 
correlation effects beyond the standard local density approximation can be simulated, 
at least partially, by means of a parameter free first principles DLM approach, 
and to see if this picture gives an adequate description of the actinides. 
The DLM picture, even if it is insufficient for a complete description, 
might lead to some insights for the understanding of the electronic structure of the actinides.

Consider, first, the status of the DLM approach in the many-body lattice
models like Hubbard or s-f exchange (``Kondo lattice'') model.
The key point is an equivalence between a many-body interacting system with Coulomb on-site interactions 
and a one-electron system in fluctuating charge and spin fields. 
This equivalence, which can be proven by the Hubbard-Stratonovich transformation,
is a base of spin-fluctuation theories of itinerant-electron magnetism \cite{moriya}.
In a complete theory, the charge and spin fields are dynamically fluctuating both in space and time. 
However, a ``static approximation'', where we neglect the dynamics of the fluctuations,\cite{hub,has}
captures an important part of the correlations while greatly simplifying the formalism, 
and may be sufficient for many problems of interest.
In this case the correlated system is described in terms of a DLM alloy 
and a CPA for this alloy becomes equivalent to the "Hubbard III" approximation \cite{hubIII} 
for the original many-body problem (see \cite{cyrot}). 
The Hubbard III approximation, and therefore also the corresponding DLM description,
gives qualitatively the correct picture of the electronic structure both in atomic and broad-band limits 
and can describe the metal-insulator (Mott) transition for half-filled bands 
when $U$ is of the order of the bandwidth \cite{hubIII}. 
The most important difference with the more elaborate DMFT picture \cite{GKKR} is an absence of the
"Kondo peak" near the Fermi level at the metallic side of the transition. 
However, this Kondo peak has a small spectral weight and, generally speaking, 
it is not obvious that it is of crucial importance for the description of 
total energy and related characteristics.
It is shown in Ref.\cite{anokhin} that the Hubbard III approximation can be 
rigorously justified, in the s-f exchange model, in the limits of large space 
dimensionality and classical spins ($S \rightarrow \infty$)
whereas for DMFT only the first limit is essential.
Note, that the Kondo resonance is essential also for a correct description
of the properties of electron states in doped Mott insulators (large
-$U$ limit near the half filling). On the other hand, for half filling
the Hartree-Fock-like descriptions of the band-split states are, in general,
quite adequate \cite{KI84}. Physically, this is the consequence of a freedom
in the description of {\it occupied} bands in terms of Bloch functions,
as well as Wannier functions or any other orthogonal linear combinations. 
The spatial fluctuations of the exchange on-site field 
can lead to a splitting of the energy spectra 
(provided the fluctuations are larger than the bandwidth) 
which corresponds, in the Hubbard model, to the Hubbard band splitting. 
The self-energy in the DLM picture is energy dependent 
and has an imaginary part describing the damping of the electrons on spin fluctuations. 
This distinguishes the DLM approach not only from magnetically ordered LSDA calculations
but also from LDA+U, SIC, and the Hartree-Fock approximation. 
In this sense, the DLM-CPA approach can be considered as a particular case of 
the ``LDA++'' approach\cite{LDA++} with a local, energy dependent, complex self-energy. 
However, in contrast to schemes taking into account Hubbard correlations, 
it is completely {\it ab initio}. A shortcoming of this approximation is an 
incorrect description of electron damping near the Fermi level; 
{\it i.e.}, the absence of the Kondo resonance, which is a consequence of 
the {\it quantum} character of spin, as well as problems describing localization
in systems that do not have half filled (or completely filled) electron shells.
However, this shortcoming may be of minor importance for 
the description of electron energy spectrum at large energy scales 
as well as for the calculations of total energy and related characteristics 
such as the equilibrium volume and elastic moduli.
Note also that this finite damping, because of electron scattering by spin fluctuations, 
is a physically correct picture for high enough temperatures.

Comparing the many-body lattice models with the density functional approach one can face
with the well-known ``Hubbard $U$ vs. Stoner $I$'' problem, i.e., with the inadequacy of the
LDA approach near the atomic limit \cite{Anisimov91,LDA++}. On the other hand, for
{\it moderately} correlated systems such as, e.g., 3d metals, the main correlation effects
are connected with spin degrees of freedom and can be described, in principle, basing on
the LSDA electronic structure. Of course, it is difficult to say {\it a priori} where the
boundary is between `moderately correlated'' and ``strongly correlated'' systems. We will
demonstrate that, at least for early actinides a LSDA-DLM description of
the electronic structure turns out to be rather successful.

Due to the localized character of the spin moments, 
any ordered magnetic structure will resemble the result of the DLM description. 
It has often been noted that localization effects could partly be accounted for 
by way of exchange spin splitting of the 5f band, 
and despite an unphysical long range magnetic ordering (absent in the DLM approach), 
several spin-polarized ferromagnetic (FM) and anti-ferromagnetic (AFM) calculations 
have been performed\cite{Skriver78,Solovyev91,Katsnelson92,Wang00,Soderlind01}.
The presented DLM model gives a natural generalization to
the paramagnetic state.

First we discuss some calculational details and then present the calculated volumes 
and bulk moduli for the actinide series. We then look at the results for $\delta$-Pu 
in detail, and the results are compared with different alternative approaches
and experiments.

\section{Calculational details}

All total energies and densities have been calculated
self-consistently within the framework of density functional theory \cite{hohenberg64,kohn65},
in the local-density approximation (LDA) in the non-magnetic cases, and within
the local-spin-density approximation (LSDA) for the spin-polarized systems
\cite{Hedin71,Barth72,ceperley80}, with the local exchange-correlation functional
by Perdew {\it et al.} \cite{PBE96}.

We used the basis set of the $s$, $p$, $d$, and $f$ linear muffin-tin orbitals
(LMTO) in the tight-binding representation and the atomic sphere approximation
(ASA) for the crystal potential
\cite{andersen75,skriver84,andersen84,andersen85}. The method was implemented
within the scalar relativistic Green's function technique
\cite{gunnarsson83,skriver91,abrikosov93b,Ruban99}. Spin-orbit coupling was not
included. The disorder of local spin moments was treated within the coherent
potential approximation (CPA), and other details relevant for the present
calculation can be found in Refs. \cite{abrikosov93b,abrikosov95}.

Since the interatomic distance is known to be the most important factor that
determines the energetic of an actinide and since we are mainly interested in
trends rather than in a detailed quantitative description of all the different
phases of the actinides, calculations were performed for the fcc crystal structure
only.

\section{Disordered local spin moment picture of the actinides}

Figure \ref{Latt} shows a comparison between experimental and calculated Wigner-Seitz radii 
for the actinide metals. For the non-magnetic LDA calculation, we find a parabolic behavior 
typical of bonding through a series such as the transition metals. 
This is in full agreement with the experimental values for the early actinides, 
but fails do describe the volumes of the heavy actinides. 
Within the DLM picture on the other hand, we find a substantial improvement.
For the light actinides, Th-U, results are identical to the non-magnetic LDA calculation, 
though for Np we find a slightly increased volume. 
For the later actinides we find an abrupt volume increase in close agreement with the experimental values 
with the transition taking place at Pu, which has an intermediate volume
close to the value of the high temperature $\delta$-phase in the fcc structure.
The results in Fig.1 are quite close to the data presented by Skriver {\it et 
al.} \cite{Skriver78}.

We find similar results for the bulk moduli.
In figure \ref{Bulk} we display a comparison between the experimental bulk modulus
for the ground state structures and the calculated bulk modulus in the fcc structure. 
The LDA calculations gives a fairly good description of the early
actinides but fails completely for the later actinides.
The DLM picture on the other hand gives a significantly improved result 
where, for example, the bulk modulus of Pu is reduced by a factor of two, 
though still somewhat higher than the experimental result for $\delta$-Pu. 

Given the fact that the results are from parameter free {\it ab initio} 
calculations without considering the exact crystal structures and spin-orbit coupling
the agreement between the calculated equilibrium volumes and bulk moduli
with experimental values is remarkably good and the results clearly indicate the ability of
the LSDA+CPA approach within the DLM picture to model important correlation
effects beyond LDA in the 5f band, without incorporating long ranged magnetic ordering.

The mechanism behind the improved description of the equilibrium volume
and bulk modulus is the formation of local disordered moments. This gives rise
to a band splitting and partial localization of the 5f electrons which reduces 
the bonding resulting in an increased lattice constant and reduced bulk modulus. 
For Th and Pa local spin moments are still quenched, and for U only a negligible 
moment of about 0.07 $\mu_{\rm B}$ is formed. 
Within our DLM scheme, neptunium has a moment of 1.69 $\mu_{\rm B}$. Though this has
little effect on the equilibrium lattice constant, the bulk modulus
is reduced by a factor of two. For Pu a spin moment of 4.76 $\mu_{\rm B}$ is
found in the DLM calculation, which corresponds to an almost complete spin polarization of the 5f electrons.
For Am and Cm a moment of 6.57 and 6.90 $\mu_{\rm B}$ is formed, respectively.
For Bk the disordered local moment is slightly reduced to 5.61 $\mu_{\rm B}$.
Notice that fully relativistic calculations, including the spin-orbit interaction,
would reduce the moments. In the case of $\delta$-Pu from about 5 $\mu_{\rm B}$ to less 
than 2 $\mu_{\rm B}$ \cite{Solovyev91} and for Am the total moment can be expected
to be $\sim$ 5 $\mu_{\rm B}$ \cite{Soderlind00}. However, the formation of a J=0
atomic ground state comes natural from a completely localized 5f shell,\cite{ReviewB} 
and is consistent with experimental data.

\section{$\delta$-Pu}

Figure \ref{Spectr} shows the calculated density of states (DOS)
for Pu in comparison with the results of photo-electron spectroscopy \cite{Arko00}. 
The LDA curve has qualitatively a very different
behavior compared to experiments. For example, the peak at the Fermi level
seen in the photoemission spectra is absent. The DLM DOS on the other hand,
is in qualitative agreement with experimental data. The DOS is
similar to the LDA-DMFT spectra by Savrasov {\it et al.} \cite{savr} (calculated at 600 K), 
but the quantitative agreement for the DLM picture is considerably better.

The electronic structure is strongly modified in the DLM picture compared 
to the LDA calculation. The 5f band is split, as shown in Fig.\ \ref{DOS},
due to the formation of local disordered moments, leading to an effective energy 
lowering of the occupied 5f band mass. The same effect is found also for the AFM 
calculation. This leads to a decreased bonding, increased equilibrium
lattice constant and a reduced bulk modulus. The effect
is similar to the results of LDA+U, LDA++, and LDA-DMFT.
In contrast to some of these models we here have an improved agreement
with the photo-electron spectra and the result is similar to
what was found in the Mixed Level Model
by Eriksson {\it et al.} \cite{Olle99}. However, though a direct full 
comparison between the Kohn-Sham eigenspectra and the photo-emission data
may not be possible, the data clearly indicates a resonance
at the Fermi level which is not present in the LDA description.

In Fig. \ref{Spectr} we also display the density of states
for an antiferromagnetic (AFM-LSDA) calculation. Though, resembling
the general DLM 5f band split, the DOS is in less agreement with 
photoemission data.

\section{discussions}

There are two principle approaches for describing the
$\alpha-\gamma$ transition in Ce and the $\alpha-\delta$ transition in Pu:
the Mott transition model by Johansson \cite{johan} 
(for Pu, it has been considered in Refs.\cite{Johansson75,Katsnelson92}) 
and the Kondo-collapse model of Allen and Martin \cite{allen}. 
In the context of DMFT, 
the appearance of the Kondo resonance is a feature of the Mott transition.
An important question is thus what is more important for the energetics of the transitions: 
the appearance of the Hubbard gap or the formation of the Kondo resonance? 
The first feature is taken into account in the DLM scheme whereas the second one is not. 
It was already mentioned that the Kondo resonance has small spectral weight, 
so it is not obvious that ignoring it is an essential shortcoming.
Moreover, both for Ce and Pu there are some arguments connecting the Mott transition 
to the peculiarities of the atomic electronic structure (atomic collapse \cite{collapse}). 
These peculiarities can obviously be taking into account in the local spin-density approach
and thus in the DLM description. Note that this collapse phenomenon, 
leading to a sharp dependence on the band width and lattice constant, 
leads to a first-order phase transition. 
Therefore the difference between the description of the Mott transition 
in the DLM or the Hubbard III approximation (continuous transition \cite{hub,anokhin}) 
and DMFT (first-order transition, see Ref.\ \cite{Kotliar02} and Refs.\ therein) 
which is very essential for the Hubbard model, 
does not play a serious role for real actinide systems where the transition is essentially of
first order for ``quasi-one-electron'' reasons. 

As for the description of magnetic properties at high temperatures $T$, 
the static approximation gives a qualitatively correct picture. 
The magnetic susceptibility $\chi$ is proportional to $<\epsilon^{2}>/T$, 
where $\epsilon$ is the exchange on-site field. 
The susceptibility corresponds to the paramagnetic Pauli spin susceptibility provided that the
exchange splitting fluctuations are Gaussian near the point $\epsilon=0$, 
and to the Curie-Weiss susceptibility provided that spontaneous spin splitting exists\cite{vons}.
The description of finite temperature magnetism of transition metals in the DLM-CPA
approach \cite{gyorffy} is different from the DMFT treatment \cite{FeNi} only by
the consideration of spins in a classical way; 
in the Heisenberg model, it corresponds to the appearance of the multiplier $S^{2}$ 
instead of the correct factor $S(S+1)$ in the Curie constant.

The DLM picture leads to an expected Curie-Weiss
behavior of the magnetic susceptibility for $\delta$-Pu. 
However, recent experiments with Al and Ga stabilized $\delta$-Pu 
shows only a vannishingly small temperature dependence of the magnetic 
susceptibility, see Fig.\ 2 in Ref.\ \cite{Meot96}. This is also
consistent with other susceptibility measurements \cite{Fournier,PuHandbook}.
The possible reason, explaining the absent Curie-Weiss behavior,
could either be due to crystal field splitting and the formation of a non-magnetic
multiplet \cite{Olle99} or because of the occurrence of a Kondo resonance that may diminish
the temperature dependence of the susceptibility provided that the resonance width 
(``Kondo temperature'') is larger than $T$. At least the first explanation
could possibly be included separately into the DLM picture, but the second one can be
described only within the framework of DMFT. 
At the same time, it is important to realize that the intention with the DLM 
picture is to model some of the main characteristics of the energetics of 
the actinides, and it does not necessarily describe the magnetic properties correctly. 

\section{summary}

In summary we have presented a first principles Disordered Local Moment (DLM) method
within the local density approximation with the disorder treated within the
coherent potential approximation. 
The DLM picture gives an reasonably good description of bond lengths 
and bulk modulus for the actinide series. 
The equivalence between the DLM picture and the Hubbard III approximation 
and their relation to the DMFT description was discussed 
and it was argued for that the DLM picture is related to DMFT through a static approximation.
The DLM density of states compares well with photoemission on $\delta$-Pu, 
in contrast to that obtained from the LDA or the magnetically ordered AFM configuration.
In general, it is found that the DLM picture gives a considerable improvement 
over the LDA results and quantitatively good agreement with experiments.

\section{Acknowledgement}

Discussions with D.\ C. Wallace, G.\ H. Landers and M. Colarieti-Tosti are gratefully acknowledged.
I.\ A.\ A. is grateful to the Swedish Research Council (VR) and the Swedish
Foundation for Strategic Research (SSF) for financial support.

\begin{figure}
\caption \small{
Comparison between the experimental \cite{ReviewB} and calculated 
atomic Wigner-Seitz radius, R$_{\rm WS}$, for the actinide metals.
\label{Latt}}
\end{figure}

\begin{figure}
\caption \small{
Comparison between the experimental and calculated bulk modulus for the
actinide metals. The experimental data are given in a) Ref. \cite{ReviewB}
and b) \cite{Duane}.
\label{Bulk}}
\end{figure}

\begin{figure}
\caption \small{
Comparison between the calculated DOS within the DLM description,
non-magnetic LDA, an antiferromagnetic (AFM) structure, and 
the photo-emission spectra (PES) \cite{Arko00} for $\delta$-Pu. 
\label{Spectr}}
\end{figure}

\begin{figure}
\caption \small{
Calculated DOS for $\delta$-Pu within the DLM description in
comparison with the non-magnetic LDA DOS and an antiferromagnetic
(AFM) LSDA DOS.
\label{DOS}}
\end{figure}


\begin{references}

\bibitem{Review} A.\ J. Freeman, G.\ H. Landers (Eds.), {\it Handbook on the Physics and
Chemistry of the Actinides}, North-Hollands, Amsterdam, 1984.
{\it Challenges in Plutonium Science}, Los Alamos Sci., {\bf 26}, 91-127 (2000).

\bibitem{Soderlind95} P. S\"oderlind, O. Eriksson, B. Johansson, J.\ M. Wills,
and A.\ M. Boring, Nature {\bf 374}, 524 (1995).

\bibitem{johan} B. Johansson, Philos. Mag. {\bf 30}, 469 (1974).

\bibitem{Skriver78} H.\ L. Skriver, O.\ K. Andersen, and B. Johansson,
Phys.\ Rev.\ Lett.\ {\bf 41}, 42 (1978).

\bibitem{ReviewB} M.\ S.\ S. Brooks, H.\ L. Skriver, and B. Johansson, in
A.\ J. Freeman, G.\ H. Landers (Eds.), {\it Handbook on the Physics and
Chemistry of the Actinides}, North-Hollands, Amsterdam, 1984.

\bibitem{Solovyev91} I.\ V. Solovyev, A.\ I. Liechtenstein, V.\ A. Gubanov,
V.\ P. Antropov, and O.\ K. Andersen,
Phys.\ Rev.\ B {\bf 43}, 14 414 (1991).

\bibitem{Soderlind00} P. S\"oderlind, R. Ahuja, O. Eriksson, B. Johansson and
J.\ M. Wills, 
Phys.\ Rev.\ B {\bf 61}, 8119 (2000).

\bibitem{Wills92} J.\ M. Wills and O. Eriksson,
Phys.\ Rev.\ B {\bf 45},  13 879 (1992).

\bibitem{Johansson75} B. Johansson,
Phys.\ Rev.\ B {\bf 11}, 2740 (1975).

\bibitem{Katsnelson92} M.\ I. Katsnelson, I.\ V. Solovyev, and A.\ V. Trefilov, 
Pis'ma ZhETF {\bf 56}, 276 (1992) [Engl. transl.: JETP Lett. {\bf 56}, 272 (1992)].

\bibitem{Ek93} J. van Ek, P.\ A. Sterne, and A. Gonis,
Phys.\ Rev.\ B {\bf 48}, 16280 (1993).

\bibitem{Soderlind94} P. S\"oderlind, O. Eriksson, B. Johansson, and J.\ M. Wills,
Phys.\ Rev.\ B. {\bf 50}, 7291 (1994).


\bibitem{Olle99} O. Eriksson, J.\ D. Becker, A.\ V. Balatsky, and J.\ M. Wills,
J.\ of Alloys and Comp., {\bf 287}, 1 (1999).

\bibitem{Wang00} Y. Wang and Y. Sun,
J.\ Phys.\ Condens. Matter {\bf 12}, L311 (2000).

\bibitem{Soderlind01} P. S\"oderlind,
Europhys.\ Lett.\ {\bf 55}, 525 (2001).

\bibitem{Anisimov91} V.\ I. Anisimov, J. Zaanen, and O.\ K. Andersen,
Phys.\ Rev.\ B {\bf 44}, 943 (1991).

\bibitem{Bouchet00} J. Bouchet, B. Siberchicot, F. Jollet, and A. Pasturel,
J.\ Phys.\ : Condens.\ Matter {\bf 12}, 1723 (2000).

\bibitem{GKKR} A. Georges, G. Kotliar, W. Krauth, and M. Rozenberg, 
Rev.\  Mod.\ Phys.\ {\bf 68}, 13 (1996).

\bibitem{anis} V.\ I. Anisimov, A.\ I. Poteryaev, M.\ A. Korotin, A.\ O. 
Anokhin, and G. Kotliar,
J.\ Phys.: Condens. Matter {\bf 9}, 7359 (1997).

\bibitem{LDA++} A.\ I. Lichtenstein and M.\ I. Katsnelson, 
Phys.\ Rev.\ B {\bf 57}, 6884 (1998).

\bibitem{savr} S.\ Y. Savrasov, G. Kotliar, and E. Abrahams, 
Nature {\bf 410}, 793 (2001).

\bibitem{LK} A.\ I. Lichtenstein and M.\ I. Katsnelson, In: {\it Band
Ferromagnetism. Ground State and Finite Temperature Phenomena}, 
ed. by K. Barbeschke, M.  Donath, and W. Nolting
(Springer, Berlin, 2001), p. 75.

\bibitem{Held} K. Held, I.\ A. Nekrasov, G. Keller, V. Eyert, N. Bluemer,
A.\ K. McMahan, R.\ T. Scalettar, T. Pruschke, V.\ I. Anisimov, and D. Vollhardt, 
In: {\it Quantum Simulations of Complex Many-Body Problems: From Theory to Algorithms}, 
ed. by J.  Grotendorst, D. Marx, and A. Muramatsu, NIC Series, Vol. 10 
(NIC Directors, Forschungzcentrum Juelich, 2002), p. 175.

\bibitem{Ce1} M.\ B. Z\"olfl, I.\ A. Nekrasov, Th. Pruschke, V.\ I. Anisimov, and J. Keller,
Phys.\ Rev.\ Lett.\ {\bf 87}, 276403 (2001).

\bibitem{Ce2} K. Held, A.\ K. McMahan, and R.\ T. Scalettar, 
Phys.\ Rev.\  Lett.\ {\bf 87}, 276404 (2001).

\bibitem{gyorffy} B.\ L. Gyorffy, A.\ J. Pindor, J. Staunton, G.\ M. Stocks
and H. Winter, J. Phys. F 15, 1337 (1985); 
J.\ B. Staunton and B.\ L. Gyorffy, 
Phys.\ Rev.\ Lett.\ {\bf 69}, 371 (1992).

\bibitem{CPA} For a review see, J.\ S. Faulkner, Prog.\ Mater.\ Sci. {\bf 27}, 1 (1982)

\bibitem{moriya} T. Moriya, {\it Spin Fluctuations in Itinerant Electron
Magnetism} (Springer, Berlin, 1985).

\bibitem{hub} J. Hubbard, 
Phys.\ Rev.\ B {\bf 19}, 2626 (1979).

\bibitem{has} H. Hasegawa, 
J.\ Phys.\ Soc.\ Japan {\bf 46}, 1504 (1979).

\bibitem{hubIII} J. Hubbard, 
Proc.\ Roy.\ Soc.\ A {\bf 281}, 401 (1964).

\bibitem{cyrot} M. Cyrot, 
Phys.\ Rev.\ Lett.\ {\bf 25}, 871 (1970).

\bibitem{anokhin} A.\ O. Anokhin, V.\ Yu. Irkhin, and M.\ I. Katsnelson, 
J.\ Phys.: Condens. Matter {\bf 3}, 1475 (1991).

\bibitem{KI84} M.\ I. Katsnelson and V.\ Yu. Irkhin,
J.\ Phys.\ C {\bf 17}, 4291 (1984).

\bibitem{hohenberg64} P. Hohenberg and W. Kohn,
Phys.\ Rev.\ {\bf 136}, B864 (1964).

\bibitem{kohn65} W. Kohn and L.\ J. Sham,
Phys.\ Rev.\ {\bf 140}, A1133 (1965).

\bibitem{Hedin71} L. Hedin and B.\ I. Lundqvist,
J.\ Phys.\ C {\bf 5}, 2064 (1971).

\bibitem{Barth72} U. von Barth and L. Hedin,
J.\ Phys.\ C {\bf 5}, 1629 (1972).

\bibitem{ceperley80} D.\ M. Ceperley and B.\ J. Alder,
Phys.\ Rev.\ Lett.\ {\bf 45}, 566 (1980).

\bibitem{PBE96} J.\ P. Perdew, K. Burke, and M. Ernzerhof,
Phys.\ Rev.\ Lett.\ {\bf 77}, 3865 (1996).

\bibitem{andersen75} O.\ K. Andersen,
Phys.\ Rev.\ B {\bf 12}, 3060 (1975).

\bibitem{skriver84} H.\ L. Skriver,
{\em The LMTO Method} (Springer-Verlag, Berlin, 1984).

\bibitem{andersen84} O.\ K. Andersen and O. Jepsen,
Phys.\ Rev.\ Lett.\ {\bf 53}, 2571 (1984).

\bibitem{andersen85}  O.\ K. Andersen, O. Jepsen, and D. Gl\"otzel,
in {\it Highlights of Condensed-Matter Theory \/}, edited by F. Bassani,
F. Fumi, and M.\ P. Tosi (North Holland, New York, 1985).

\bibitem{gunnarsson83} O. Gunnarsson, O. Jepsen, and O.\ K. Andersen,
Phys.\ Rev.\ B {\bf 27}, 7144 (1983)

\bibitem{skriver91}  H.\ L. Skriver and N.\ M. Rosengaard, 
Phys. Rev. B, {\bf 43}, 9538 (1991).

\bibitem{abrikosov93b} I.\ A. Abrikosov and H.\ L. Skriver,
Phys.\ Rev.\ B {\bf 47}, 16532 (1993).

\bibitem{Ruban99}  A.\ V. Ruban and H.\ L. Skriver, 
Compt. Mat. Sci., {\bf 15}, 119 (1999).

\bibitem{abrikosov95} A.\ I. Abrikosov,  O. Eriksson, P. S\"oderlind, 
H.\ L. Skriver, and B. Johansson, 
Phys.\ Rev.\ B {\bf 51}, 1058 (1995).

\bibitem{Arko00} A.\ J. Arko, J.\ J. Morales, 
J. Wills, J. Lashley, F. Wastin, and J. Rebizant, 
Phys.\ Rev.\ B {\bf 62}, 1773 (2000).

\bibitem{allen}J. W. Allen and R.\ M. Martin, 
Phys.\ Rev.\ Lett.\ {\bf 49}, 1106 (1982).

\bibitem{collapse} V.\ V. Kamyshenko, M.\ I. Katsnelson, A.\ I.  Lichtenstein, 
and A.\ V. Trefilov,
Fiz. Tverdogo Tela {\bf 29}, 3581 (1987).

\bibitem{Kotliar02} G. Kotliar, 
J.\ Low Temp.\ Phys.\ {\bf 126}, 1009 (2002).

\bibitem{vons} S.\ V. Vonsovsky and M.\ I. Katsnelson, Physica B {\bf
159}, 61 (1989).

\bibitem{FeNi} A.\ I. Lichtenstein, M.\ I. Katsnelson, and G. Kotliar,
Phys.\ Rev.\ Lett.\ {\bf 87}, 067205 (2001).

\bibitem{Duane} D.\ C. Wallace,
Phys.\ Rev.\ B {\bf 58}, 15 433 (1998).

\bibitem{Meot96} S.\ Meot-Reymond, and J.\ M.\ Fourier,
J.\ Alloys and Comp. {\bf 232}, 119 (1996).

\bibitem{Fournier} J.\ -M. Fournier and R. Troc, in
A.\ J. Freeman, G.\ H. Landers (Eds.), {\it Handbook on the Physics and
Chemistry of the Actinides}, North-Hollands, Amsterdam, 1984.

\bibitem{PuHandbook} in O.\ J. Wick (Editor), {\it Plutonium Handbook},
American Nuclear Society, 1980.


\end{references}
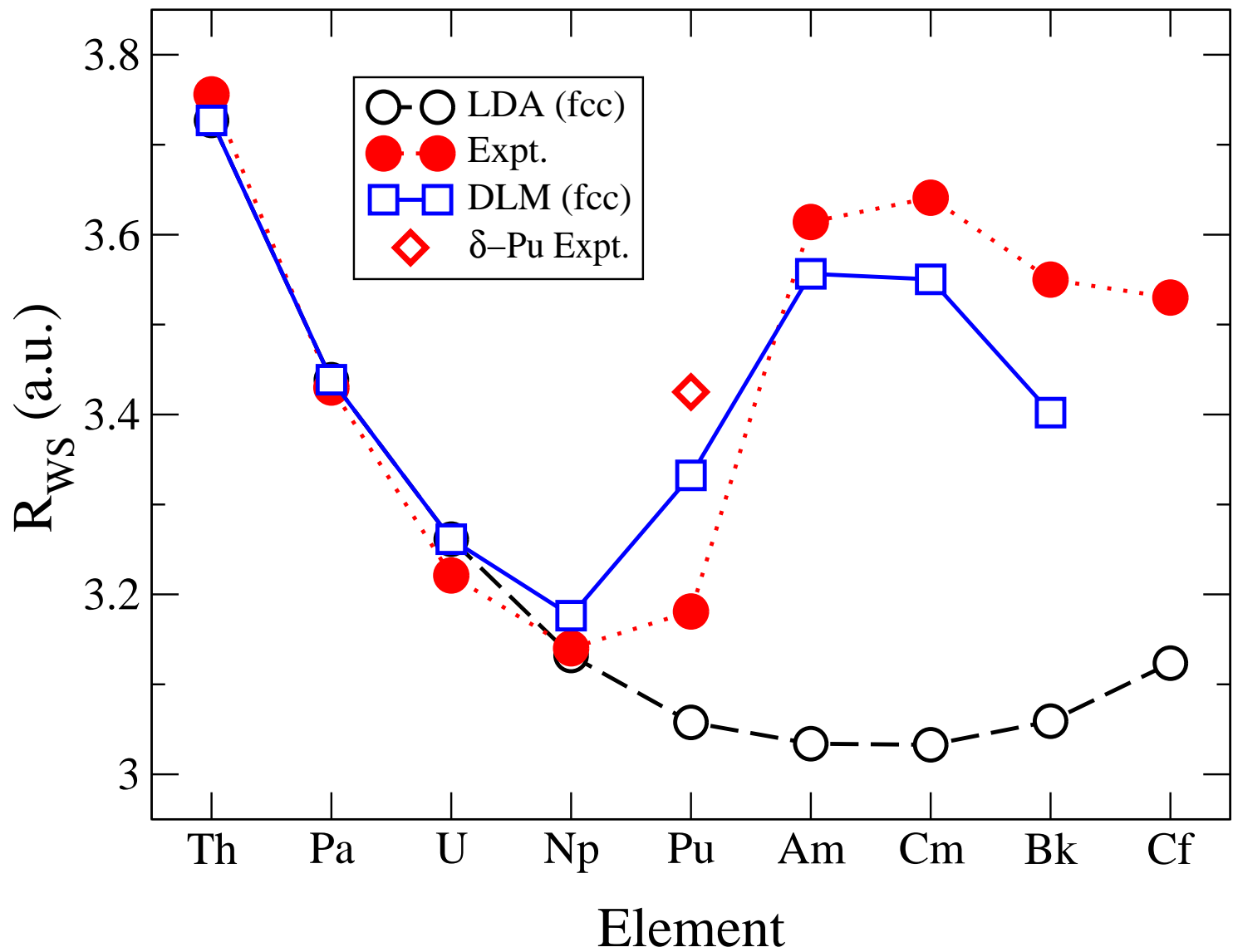
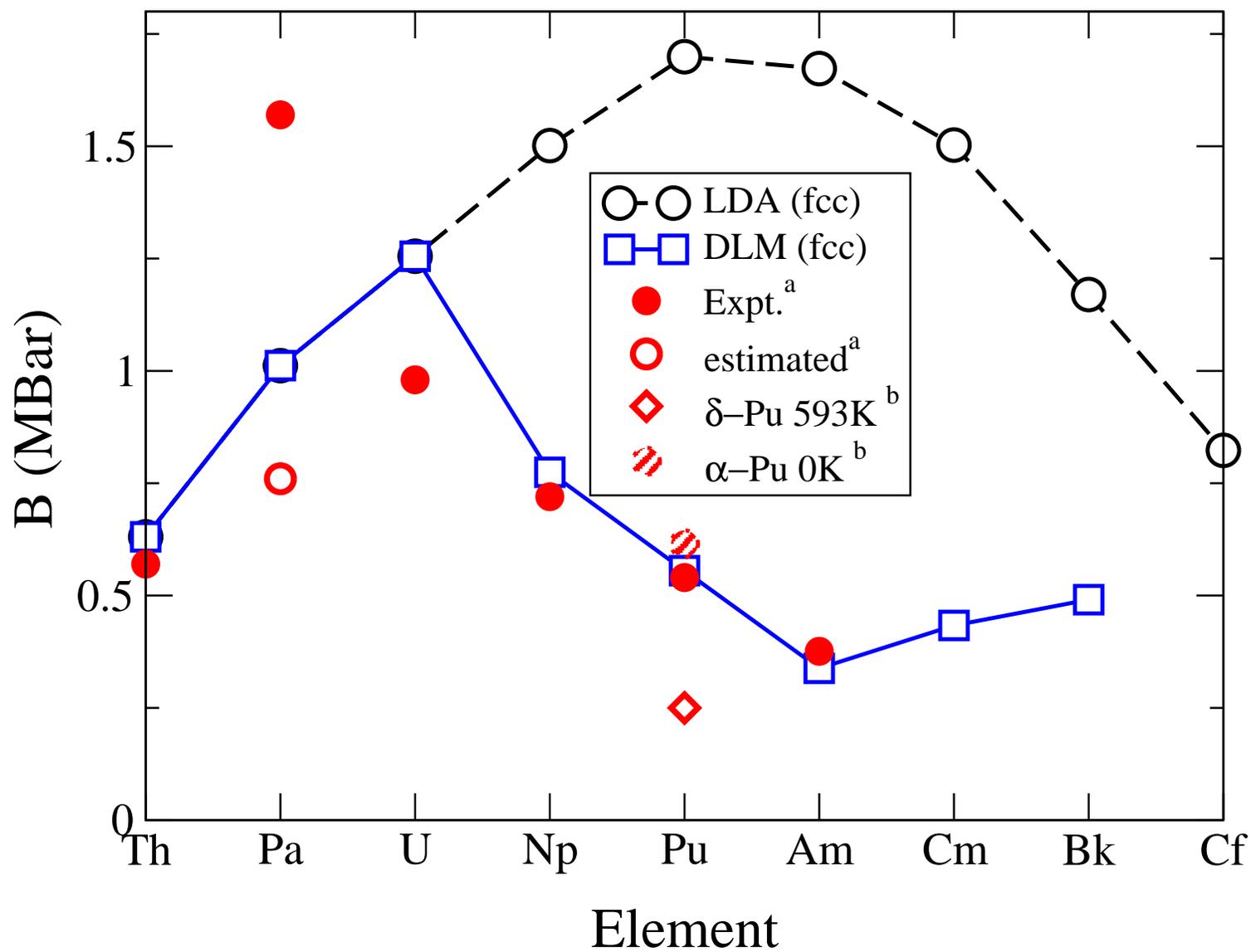
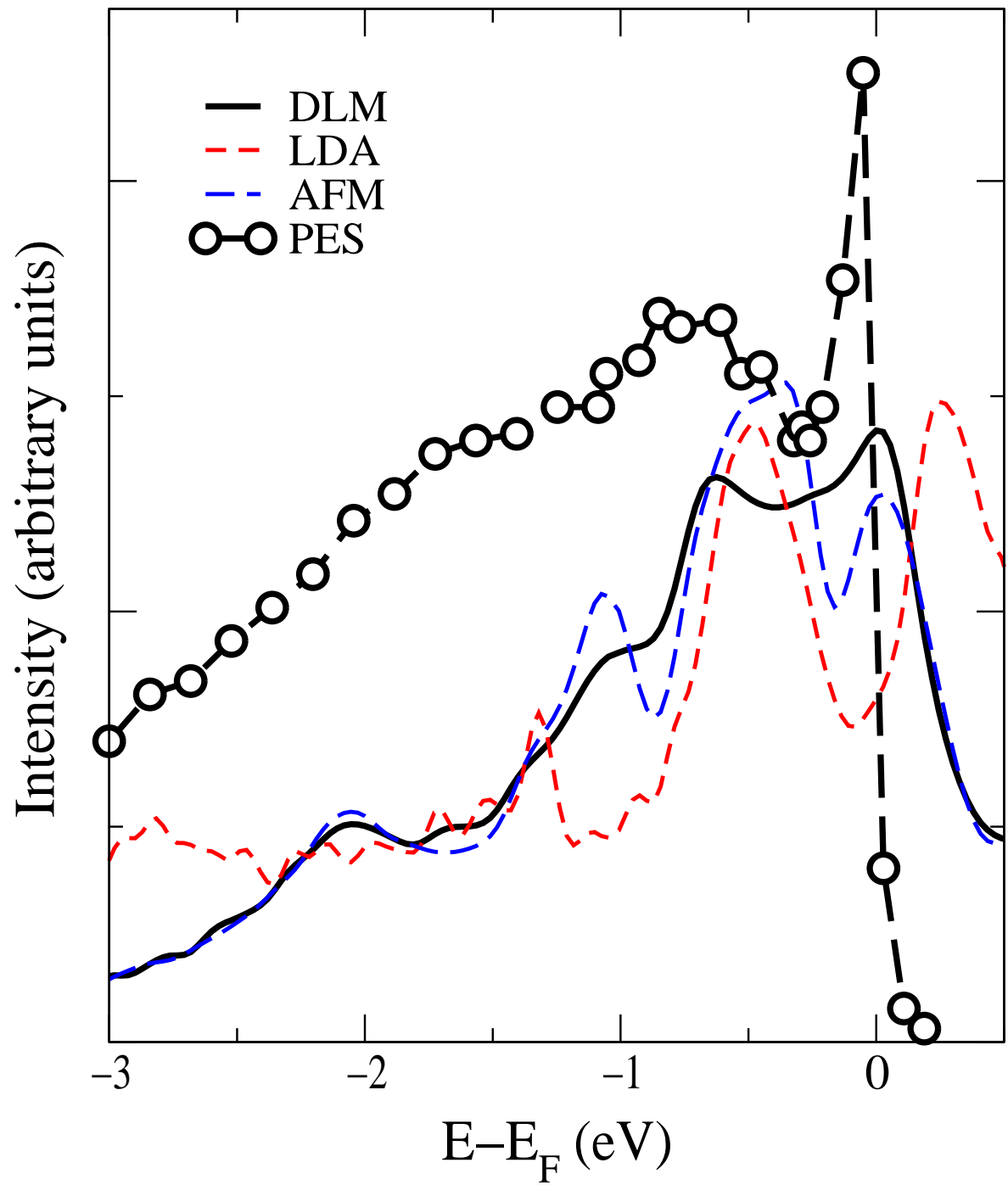
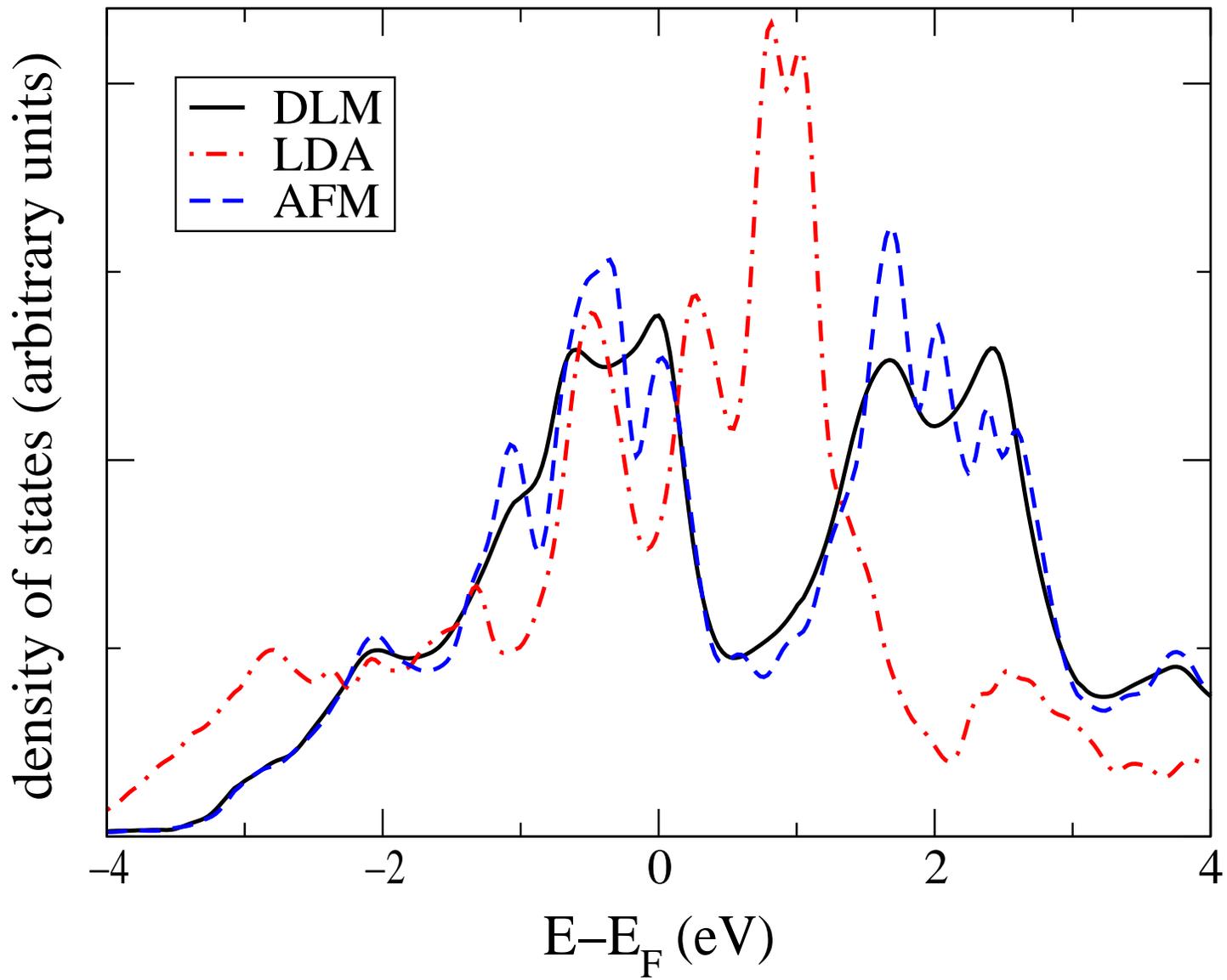
\end{document}